>1# InGaAsP/InP uni-travelling-carrier photodiode at 1064nm wavelength

>Zhiyang Xie, Yaojiang Chen, Ningtao Zhang and Baile Chenabstract>*Abstract*—High-speed back-illuminated uni-traveling-carrier photodiodes (PDs) at 1064nm were demonstrated grown on InP with 3-dB bandwidth of 17.8 GHz at -5 V bias, using InGaAsP as absorption layer. PDs with 40 *μ*m-diameter deliver RF output power levels as high as 19.5 dBm at 13 GHz. This structure can achieve low dark current density of $1\times10^{-8}$ A/cm$^2$ at $-5$ V bias and quantum efficiency of 45.2% at 1064nm. An analytical model based on S-parameter fitting was built to extract parameter to assess the bandwidth limiting factors.

*Index Terms*—UTC photodiode, 1064nm wavelength, low dark current, high-speed## I. INTRODUCTION

In recent years, the application of the 1064nm wavelength band has been widely aroused in the laser ranging, altimetry, nonlinear optics, medical treatment, coherent detection, and atmospheric remote sensing, which benefits from the outstanding features of Nd:YAG lasers with characteristic of high gain, low threshold and good optical properties [1, 2]. Yb-doped fiber amplifiers (YDFA's) also give the potential to build RF photonic system, offering broadband amplification (from 975 nm to 1200 nm), excellent power conversion efficiency and high gain in a short length of fiber [3, 4]. On the other hand, YDFA's can avoid excited state absorption and concentration quenching by interionic energy transfer, which become well-known complications for erbium-doped amplifiers(EDFA's) [3]. High-power PDs with large bandwidth are key components to RF photonic system [5-7], providing the potential to eliminate electronic power amplifiers and replace optical devices in coherent detection system. Recently, photodiodes with high quantum efficiency of 98%±0.8% at 1064nm has been presented in reference [8]. However, there is few researches on high-speed PDs at this wavelength. The bandwidth and RF power of uni-traveling-carrier (UTC) PDs have been demonstrated to be higher than p-i-n PDs [9], since only photo-generated electrons are injected into the drift layer and give rise to the photocurrent, while the excess hole quickly decay at the dielectric relaxation time in p-doped absorber.

>This work was supported in part by the Shanghai Sailing Program under Grant 17YF1429300, in part by the ShanghaiTech University startup funding under Grant F-0203-16-002 (Corresponding authors: Baile Chen.)

Z. Xie, Y.Chen, N.Zhang and B. Chen was with School of Information Science and Technology, Shanghaitech University, Shanghai, SH 021 China. (e-mail: xiezhy@shanghaitech.edu.cn; chenyj@shanghaitech.edu.cn; zhangnt@shanghaitech.edu.cn; chenbl@shanghaitech.edu.cn)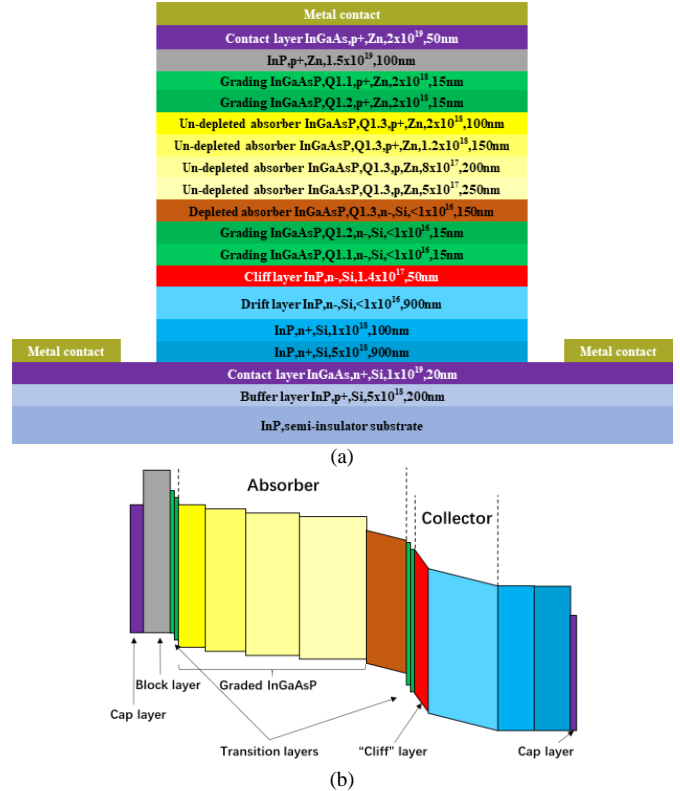

Fig. 1. (a) Epitaxial layer design. (b) band diagram of InGaAsP/InP UTC PDs.

In this letter, we report high-speed back-illuminated UTC PDs with InGaAsP absorption material grown on InP for 1064nm high speed detection, which show 3-dB bandwidth of 17.8 GHz at -5 V bias. PDs with 40 *μ*m-diameter deliver RF output power levels as high as 19.5 dBm at 13 GHz. It is also noted that PDs with InGaAs absorber can also be used for 1064nm detection. However, InGaAsP materials have larger bandgap than the InGaAs, which are expected to have less thermal generated rate and thus less dark current. Therefore, it should be desirable to design the PD with larger bandgap absorber while targeting at same wavelength. The detailed epitaxial design of device, measurement setup, and experimental results are described. Bandwidth limiting factors were analyzed through a model that utilized the parameters extracted from S11 fitting.

## II. DEVICE DESIGN

The epitaxial layer design of PDs and band diagram are shown in Fig. 1. The structure was grown on semi-insulating

InP substrates. A 100 nm n-type InP layer with a doping concentration of $1\times10^{18}$ cm$^{-3}$ is then grown to reduce Si diffusion into drift layer [10]. A 900 nm-thick lightly n-doped InP electron drift layer was designed to reduce the junction capacitance and thus increased the RC-bandwidth limitation. A 50 nm-thick n-doped ($1.4\times10^{17}$ cm$^{-3}$) InP cliff layer was used to maintain high electric field in the depleted absorber [11]. Two 15 nm-thick p-doped InGaAsP quaternary layers with larger bandgap were used to smooth the conduction band offset between InP and InGaAsP absorber, which can suppress charge accumulation at the heterojunction interfaces [12]. In p-type absorbing layer, the doping of InGaAsP was graded in four steps in order to create a quasi-electric field that assists electron transport [12]. The p-type contact was formed using 100 nm heavily p-doped InP and 50 nm InGaAs.

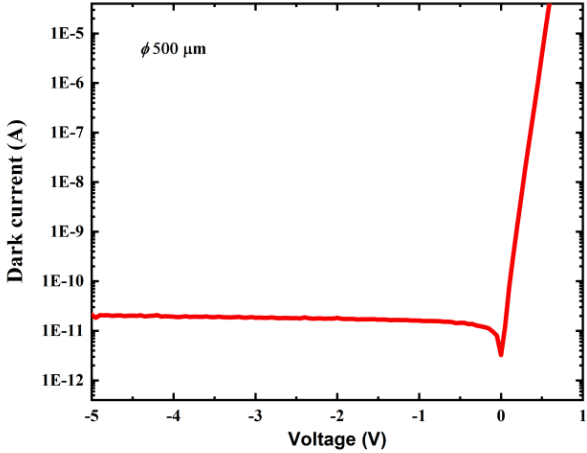

Fig. 2. Dark current versus voltage for PDs with diameters of 500 $\mu$m.

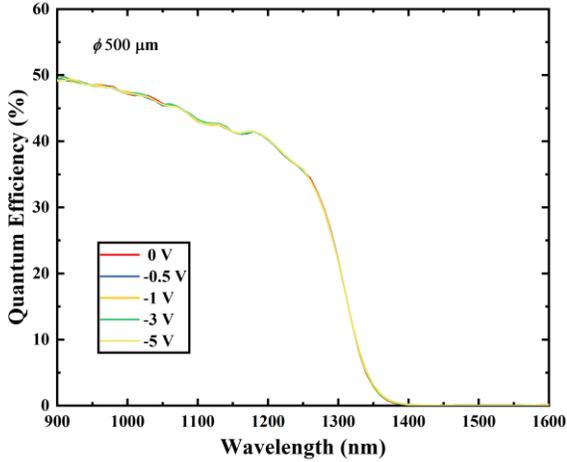

Fig. 3. Quantum efficiency of the device with diameter of 500 $\mu$m at different bias voltage.

As the excess holes in the p-doped absorption layer decay quickly and only electrons (with higher mobility than hole) transit the drift layer, UTC PDs perform higher bandwidth than traditional p-i-n PDs. The space charge accumulation at high RF output, which is referred to as the space charge effect, can also be suppressed by this uni-travelling-carrier effect associated with the p-doped cliff layer [11]. UTC PDs were fabricated by a double mesa process through lithography and wet etch which is finally connected to gold-plated bonding pads through an air-bridge.

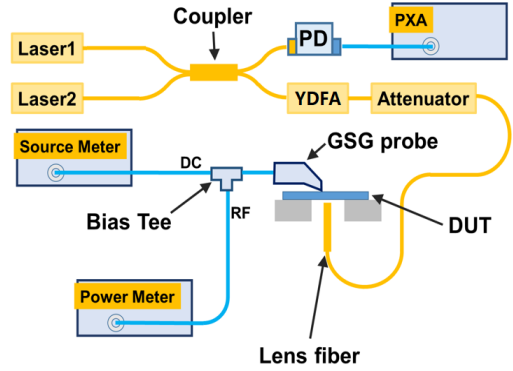

Fig. 4. Schematic of optical heterodyne setup for bandwidth and saturation characterization

### III. CHARACTERIZATION

#### A. Dark current and quantum efficiency

The device with 500 $\mu$m diameters was fabricated to measure the dark current and quantum efficiency. The dark current versus bias voltage characteristics are shown in Fig. 2. The PDs show low dark current of $2\times10^{-11}$ A at $-5$ V. The corresponding dark current density is $1\times10^{-8}$ A/cm$^2$, which is near four order lower than those of UTC-PDs with InGaAs absorber reported in Ref [13-15].

The quantum efficiency was measured from 900nm to 1600nm by comparing the photo response of the InGaAsP UTC device to that of a calibrated commerical Ge photodiodes. A tungsten lamp with broadband spectrum followed by a grating spectrometer was used to generate a monochromatic light. In order remove the high order short wavelength generated in the spectrometer, a longpass filter with cuffoff wavelength of 830nm was used during the measurement.

The quantum efficiency of the device without anti-reflection coating at the wavelength range 900 nm to 1600 nm is shown in Fig. 3, where the typical quantum efficiency of the PDs at 1064nm achieves 45.2% at 0 V. The device exhibites very similar performance at different bias from 0 V to -5 V, indicating the long diffusion length of the InGaAsP absorber.

#### B. Bandwidth

The Optical heterodyne setup is showed in Fig. 4. Two distributed feedback (DFB) lasers with wavelengths near 1064nm were heterodyned to generate a modulated optical signal. An electric spectrum analyzer along with reference PD was used to track the beat frequency from one output arm of the optical fiber coupler, which was controlled by thermally tuning the one of the DFB lasers. The amplitude of optical signal was controlled by an YDFA at the other output arm of the coupler, which can provide stable and high power beam, together with a variable optical attenuator, aimed at maintaining different levels of light intensity in bandwidth measuement setup. A lensed fiber was used to coupled light into the back-illuminated device. The RF power was finally measured by a R&S power

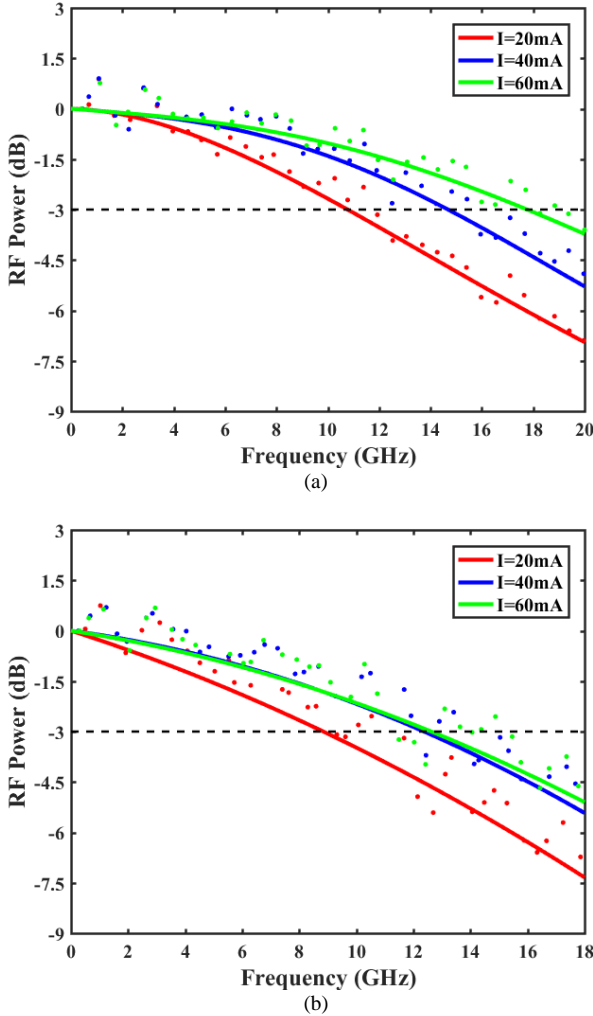

Fig. 5. Frequency response at -5 V bias, (a): 34 $\mu$m-diameter device (b): 40 $\mu$m-diameter device (solid lines are polynomial fitting curves)

where the responsivity dropped by 50% during the measurement.

Fig. 5 shows the result of measured 3-dB bandwidth of PDs with 34 and 40 $\mu$m-diameter at 20, 40 and 60 mA average photocurrent and 5 V reversed bias with 50 Ω load. The 34 $\mu$m-diameter device achieved 17.8 GHz with photocurrent of 60 mA at -5 V. It can be observed that bandwidth is enhanced with higher photocurrent, which is attributed to the fact that the photo-generated carriers induce an electric field assisting electron transport in the un-depleted absorption region.

*C. RF Power Saturation*

The setup of RF power saturation measurement is similar to the bandwidth measurement as described above in Fig. 5. Saturation is characterized by measuring the RF output as a function of average photocurrent at a fixed frequency. Fig. 6 show the RF output power of 34 and 40 $\mu$m- diameter devices at 3-dB bandwidth and -5 V. PDs with 40 $\mu$m-diameter can deliver high RF output power levels up to 19.5 dBm at 13.0

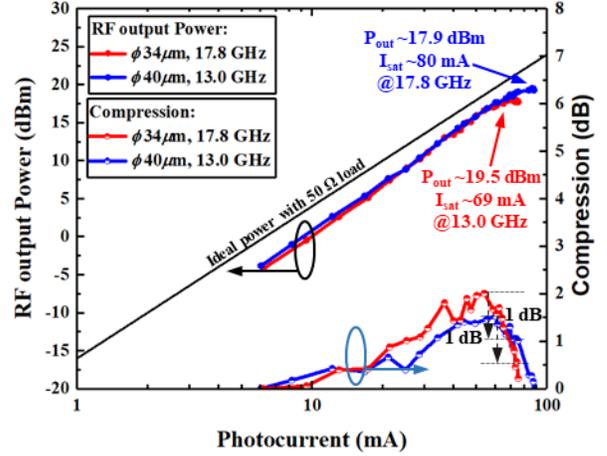

Fig. 6. RF power versus average photocurrent for 34 and 40 $\mu$m-diameter devices at 3dB-bandwidth and -5 V bias

GHz, and the device with 34 $\mu$m-diameter can reach 17.9 dBm at 17.8 GHz. It is observed that the power increases prominently with increasing photocurrent before it saturates, which indicates the enhancement of bandwidth at higher photocurrent. The saturation current is defined as the average photocurrent where the RF power compression curve drops by 1 dB from its peak value. The 40 $\mu$m-diameter device saturates at 80 mA photocurrent at 13.0 GHz. And the 34 $\mu$m-diameter device exhibits lower saturation photocurrent of 69 mA at 17.8GHz.

*D. S11 fitting and bandwidth limiting factors*

The scattering parameter S11 of back-illuminated device was measured by using a Vector Network Analyzer (VNA) up to 67 GHz. The parameter fitting was done in Advanced Design System (ADS) software, with equivalent circuit model is shown in Fig. 7. $I_s$, $R_p$, $C_{pn}$, $R_s$ represent the AC current source, the junction resistance, the junction capacity and the series resistance, respectively. The 3D-model of gold-plated pad and air-bridge was built and extracted from Ansys electromagnetic simulation software, which involves the parasitic capacitance and parasitic inductance of the air-bridge.

Fig. 8 shows the measured S11 on Smith charts and the fitting results for PDs with 40 $\mu$m-diameter and 34 $\mu$m-diameter at the reversed bias of 5 V. The extracted parameters are summarized in Table I, as well as the capacitances calculated by Eq. (1).

$$C = \frac{\varepsilon_0 \varepsilon_r A}{d} \quad (1)$$

where $\varepsilon_0$, $\varepsilon_r$, $A$, $d$ are the permittivity of free space, the dielectric constant of InP, area of device and width of depletion region, respectively. The extracted $R_p$, typically on the order of Gig-Ω, and parasitic capacitance and parasitic inductance of air-bridge, which are included in the 3D-model of gold-plated pad in HFSS software, are not shown in this table. Using the parameter extracted from S11 fitting, the overall 3-dB bandwidth can be estimated by [16]:

$$f_{3dB} = \sqrt{\frac{1}{f_{tr}^2 + f_{RC}^2}} \quad (2)$$

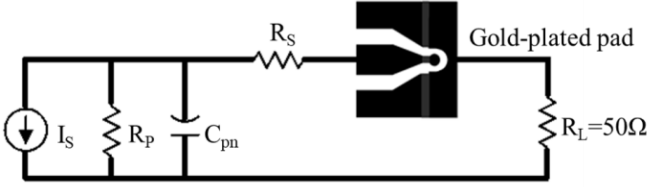

Fig. 7. UTC PDs circuit model for S11 fitting.

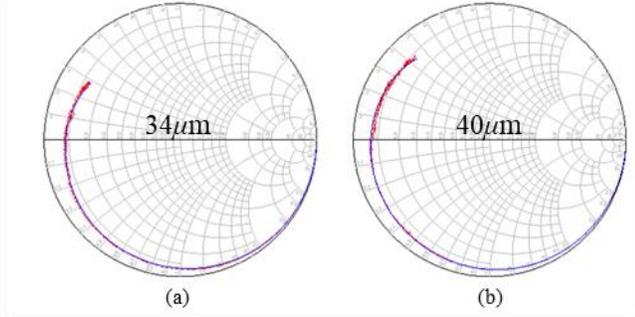

Fig. 8. Measured and fitted (smooth line) S11 data with 10 MHz-67 GHz frequency range at -5 V bias for UTC PDs of (a) 34 μm-diameter, (b) 40 μm-diameter.

TABLE I
EXTRACTED AND CALCULATED PARAMETERS.

| Extracted from S11 fitting at -5 V | | | Calculated by Eq. (1) |
|---|---|---|---|
| Diameter (μm) | $C_{pn}$ (fF) | $R_s$ (Ω) | $C_{pn}$ (fF) |
| 34 | 85 | 5 | 98 |
| 40 | 116 | 4.5 | 135 |

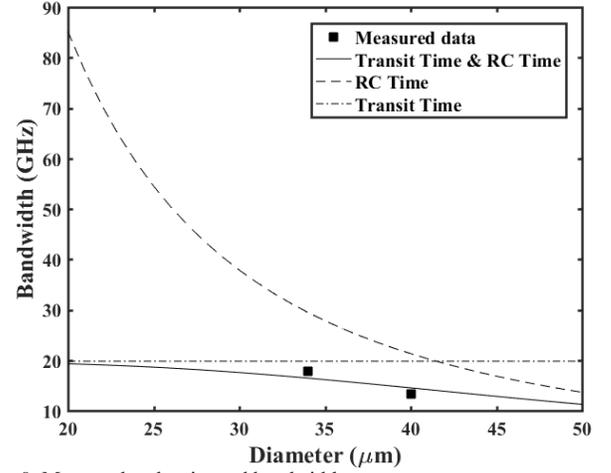

Fig. 9. Measured and estimated bandwidth

where $f_{tr}$ and $f_{RC}$ denote the transit-time-limited and RC-limited bandwidth. The transit-time-limited bandwidth can be calculated through the time accumulated by an electron that transporting in un-depleted InGaAsP absorber and depleted InP drift layer.

Fig. 9 shows good agreement between the simulated and measured 3-dB bandwidth. It is also indicated that the device with diameter less than about 40 μm is transit-time-limited for this epitaxial design. However, it is also noted that the total 3-dB bandwidth is still under the influence of the device size as the diameters of the device under test (34 and 40 μm-) are close to this critical size, which can be interpreted by the Eq. (2).

## IV. CONCLUSION

In this letter, we reported back-illuminated UTC PDs with 3-dB bandwidth of 17.8 GHz at 1064nm wavelength. Under -5 V bias, PDs with 40 μm-diameter can deliver RF output power level as high as 19.5 dBm at 13 GHz, with saturation current of 80 mA. The structure can achieve low dark current density of $1 \times 10^{-8}$ A/cm$^2$ at $-5$ V bias and quantum efficiency of 45.2% at 1064nm wavelength. An analytical model was built to fit the S-parameter in order to assess the bandwidth factor of this epitaxial design. It is found that the bandwidth of the device with diameter less than 40 μm was mainly limited by carrier transit time in PD.

REFERENCES

[1] M. Ross, "YAG laser operation by semiconductor laser pumping," *Proceedings of the IEEE,* vol. 56, no. 2, pp. 196-197, Feb. 1968.
[2] B. Zhou, T. J. Kane, G. J. Dixon, and R. L. Byer, "Efficient, frequency-stable laser-diode-pumped Nd:YAG laser," *Opt. Lett.,* vol. 10, no. 2, pp. 62-64, Feb. 1985.
[3] R. Paschotta, J. Nilsson, A. C. Tropper, and D. C. Hanna, "Ytterbium-Doped Fiber Amplifiers," *IEEE Journal of Quantum Electronics,* vol. 33, no. 7, pp. 1049-1056, July 1997.
[4] R. I. Laming, S. B. Poole, and E. J. Tarbox, "Pump excited-state absorption in erbium doped fiber," *Opt. Lett.,* vol. 13, no. 12, pp. 1084-1086, 1988.
[5] T. Nagatsuma, H. Ito, and T. Ishibashi, "High-power RF photodiodes and their applications," *Laser & Photonics Review,* vol. 3, no. 1-2, pp. 123-137, 2009.
[6] T. M. Fortier *et al.*, "Photonic microwave generation with high-power photodiodes," *Opt Lett,* vol. 38, no. 10, pp. 1712-4, May 15 2013.
[7] A. Beling *et al.*, "High-Power Photodiodes for Analog Applications," *Microwave Photonics (MWP) and the 2014 9th Asia-Pacific Microwave Photonics Conference (APMP) 2014 International Topical Meeting on,* pp. 412-414, 2014.
[8] J. Zang *et al.*, "High Quantum Efficiency Uni-Traveling-Carrier Photodiode," *IEEE Photonics Technology Letters,* vol. 29, no. 3, pp. 302-305, 2017.
[9] T. Ishibashi, N. Shimizu, S. Kodama, H. Ito, T. Nagatsuma, and T. Furuta, "Uni-traveling-carrier photodiodes," in *Proc. Tech. Dig. Ultrafast Electron. Optoelectron. OSA Spring TopicalMeeting*, 1997, vol. 13, pp. 166-168.
[10] P. Huapu, L. Zhi, and J. C. Campbell, "High-Power High-Responsivity Modified Uni-Traveling-Carrier Photodiode Used as V-Band Optoelectronic Mixers," *Journal of Lightwave Technology,* vol. 28, no. 8, pp. 1184-1189, 2010.
[11] H. Pan, A. Beling, H. Chen, and J. C. Campbell, "Characterization and optimization of high-power InGaAs/InP photodiodes," *Optical and Quantum Electronics,* vol. 40, no. 1, pp. 41-46, 2008.
[12] Z. Li, H. Pan, H. Chen, A. Beling, and J. C. Campbell, "High-Saturation-Current Modified Uni-Traveling-Carrier Photodiode With Cliff Layer," *IEEE Journal of Quantum Electronics,* vol. 46, no. 5, pp. 626-632, 2010.
[13] Q. Li *et al.*, "High-Power Flip-Chip Bonded Photodiode With 110 GHz Bandwidth," *Journal of Lightwave Technology,* vol. 34, no. 9, pp. 2139-2144, 2016.
[14] Q. Meng, H. Wang, C. Liu, X. Guo, J. Gao, and K. S. Ang, "High-Speed and High-Responsivity InP-Based Uni-Traveling-Carrier Photodiodes," *IEEE Journal of the Electron Devices Society,* vol. 5, no. 1, pp. 40-44, 2017.
[15] Q. Li *et al.*, "Waveguide-Integrated High-Speed and High-Power Photodiode with >105 GHz Bandwidth," presented at the 2017 IEEE Photonics Conference (IPC), Orlando, FL, 2017.
[16] K. Kato, "Ultrawide-band_high-frequency photodetectors," *IEEE Transactions on Microwave Theory and Techniques,* vol. 47, no. 7, pp. 1265-1281, Jul 1999.